# Many-body origin of the "trion line"

M. COMBESCOT, J. TRIBOLLET, G. KARCZEWSKI*, F. BERNARDOT,

C. TESTELIN and M. CHAMARRO

*Groupe de Physique des Solides - Universités Paris-VI et Paris-VII, CNRS UMR 75 88,*

*2 place Jussieu, 75251 Paris Cedex 05, France*

*\*Institute of Physics, Polish Academy of Sciences, Al. Lotnikow 32/46, 02-668 Warsaw, Poland*

## Abstract

We show that the so-called "trion line" in the absorption spectrum of doped quantum wells, comes from a singular many-body object, *intrinsically wide in energy*: the photocreated virtual exciton dressed by Coulomb *and* Pauli interactions with the well carriers. This understanding is supported by the spectra of circular dichroism obtained with a spin-polarized Fermi sea: the sharp edge on the low-energy side and the significant tail at high energies are well explained by these many-body effects, *not* by bound 3-body trions.

Although the existence of charged excitons, called trions, has been predicted long ago [1], experimental evidence of states below the exciton one was achieved recently only. The exciton being neutral, its carrier attraction is very weak so that the trion binding energy has to be very small. However, as all binding energies are increased by a reduction of dimensionality, it was expected these trions to be possibly seen in doped quantum wells [2,3]. This is why it appears as natural to assign the line below the exciton one, observed in these wells, to the formation of trions.

Since then, many experimental works [4-13] have been devoted to the study of this new "trion line". Measurements of lifetimes, polarization and screening effects, as well as the evolution of the oscillator strength of the line with doping, have been explained rather satisfactorily by independent photo-created trions, although it was felt that they should strongly interact with the electron gas.

In this letter, we show why this usual explanation is too simple. Our new experimental data support the idea that the photon absorption in doped quantum wells does *not* generate a 3-body object, the trion, but a many-body object which results from Coulomb and Pauli interactions between the suddenly created virtual exciton and the carriers present in the sample: The spectrum of circular dichroism in the presence of a polarized electron gas, does show a sharp edge at low energy and a long tail at high energy, which are well explained by these many-body effects but not by a set of trions.

The main characteristic of the "trion line" is its strong increase with doping. This may appear as somewhat obvious: because one carrier has to be added to the photocreated exciton to form a trion, there is no hope to form any trion if the well has no carrier. Moreover, the more carriers, the more trions can form: Because of this, one is pushed to think that the "trion line" has to increase with doping. This way of thinking is however questionned by the following remarks:

(i) Trions are composite fermions; there is no way to pile them up all at the *same* energy, due to some kind of underlying Pauli exclusion.

(ii) The absorption line, measured in the weak probe limit, shows the *linear* response of the sample. In a linear response, only one electron-hole pair is photocreated by construction. With one hole, there is no way to make more than one trion, whatever the number of available carriers.

(iii) We have recently shown that the trion oscillator strength [14] is one trion volume divided by one sample volume, smaller than the exciton oscillator strength, making the trion impossible to see in the infinite sample limit.

Consequently, the low-energy line in the absorption spectra of doped quantum wells can hardly be due to a trion. We now discuss what physically happens when a photon tuned well below the conduction band, is absorbed by a doped well, first in terms of trions – to show the inconsistency – and then in terms of a virtual exciton interacting with carriers.

*a) Let us start thinking in terms of trions.*

An impinging photon of momentum $\mathbf{Q_p}$ with a Fermi sea electron of momentum $\mathbf{k}$, can create a 3-body object with a center of mass momentum $\mathbf{k} + \mathbf{Q_p} \approx \mathbf{k}$. The photon energy, relative to the energy gap, necessary to form such a trion is $\Delta_{\mathbf{k}}$ with $\Delta_{\mathbf{k}} + \hbar^2 \mathbf{k}^2/2m_e = \varepsilon_{\eta_0} + \hbar^2(\mathbf{k}+\mathbf{Q_p})^2/2(2m_e + m_h)$, where $\varepsilon_{\eta_0}$ is the trion relative motion ground state energy. This gives a delta peak in the absorption spectrum, at $\omega = \Delta_{\mathbf{k}}$. Since this peak moves to low energies when $\mathbf{k}$ increases, a Fermi sea of electrons with $\mathbf{k}$ between ($\mathbf{0}$, $\mathbf{k_F}$), would give rise, within this trion picture, to a constant absorption between $\Delta_{\mathbf{k_F}}$ and $\Delta_{\mathbf{k=0}}$, see Fig.2(a) – because the 2D density of states is constant while the trion oscillator strength is almost constant if $\mathbf{k}$ stays small –. For finite temperature, this absorption would have a low-energy tail, as electrons above $\mathbf{k_F}$ then exist [15].

Actually, these simple trions do not hold out against many-body effects: Indeed, the remaining Fermi sea electrons feel the trion made with the $\mathbf{k}$ electron and react, either through Pauli exclusion, *i.e.*, carriers exchange, or through Coulomb interactions, *i.e.*, excitations of electron-hole pairs across the Fermi level or inside the sea to fill up the hole left by the $\mathbf{k}$ electron (see Fig.1(a)). The first type of excitations, having a positive energy, produces a high-energy tail to the $\Delta_{\mathbf{k}}$ peak, while the second type, which can have a negative energy if the electron which fills the hole is above it, adds a low-energy tail (ending at $\Delta_{\mathbf{k_F}}$). Consequently, the bound trion with a well-defined energy, associated to the $\Delta_{\mathbf{k}}$ peak, is completely washed out by these pair excitations.

Among the Fermi sea electrons, the ones at the Fermi level are however somewhat peculiar because the hole they leave, is at the top of the sea, so that there is no pair excitation with negative energy to fill it: the delta peak at $\omega = \Delta_{\mathbf{k_F}}$ only has a high-energy tail.

This leads us to think that, due to many-body effects, the low-energy line of the absorption spectrum of a doped well, must be intrinsically broad, independently from any additional extrinsic broadenings. This broad line must start at $\Delta_{\mathbf{k_F}}$ (if we forget the Fumi shift) and extend well above $\Delta_{\mathbf{k=0}}$, due to excitations of electron-hole pairs in the Fermi sea. This broad line should moreover have a sharp edge on its low-energy side, most probably singular, due to similarities with Fermi edge singularities (see Fig.2(c)).

To support these ideas, we have measured the circular dichroism in a doped quantum well when the Fermi sea is polarized. This dichroism shows the difference in the absorption of $\sigma_\pm$ photons. If the Fermi level for (-1/2) electrons is above the one for (+1/2) electrons, as in Fig.1(b), the absorption spectrum of $\sigma_-$ photons must start below the one of $\sigma_+$ photons. If trions were formed, we should have the constant absorptions shown in Fig.2(a), while, if many-body effects are included, we must have absorptions with sharp edges on their low-energy sides, and long tails at high energy, as shown in Fig.2(c). Their differences should thus look like the curves of Fig.2(b, d). The measured circular dichroism, obtained with the experimental set-up described below (see Fig.3(b)), is in good qualitative agreement with curve (d) while it rules out curve (b), *i. e.*, a description with simple trions. As already mentioned, a finite temperature would induce a low-energy tail, not the high-energy tail experimentally seen.

*b) Let us now think in terms of a photocreated virtual exciton interacting with carriers.*

In a quantum well, a $\sigma_+$ photon creates a virtual exciton made with a (+3/2) hole and a (–1/2) electron. The linear response measured in a probe absorption (quadratic in the semiconductor-photon coupling), results from all possible interactions between this virtual exciton and the Fermi sea, leaving this Fermi sea unchanged after the exciton recombination (see Fig.4).

One possibility is to have a set of *direct Coulomb scatterings* [16] between the exciton and the carriers. The exciton is then always made with the same electron; so that the hole finally recombines with the photocreated electron. The simplest of these direct Coulomb processes, in which one Fermi sea electron only is involved, is shown in Fig.4(a). These direct Coulomb scatterings exist with both ($s=\pm 1/2$) electrons, whatever the photocreated electron spin, so that they do not differentiate $\sigma_\pm$ photons.

Due to the exciton composite nature [16], another possible interaction is a carrier exchange between the photocreated virtual exciton and the Fermi sea, through Pauli scatterings. These exchanges can exist without Coulomb process, as in Fig.4(b), or mixed with them, as in Fig.4(c). The intermediate exciton is then made with any Fermi sea electron with spin ($s = \pm 1/2$). However as, in a quantum well, a (+3/2) hole can only recombine with a (-1/2) electron, the processes in which the hole finally recombines with an electron different from the photocreated one, are more numerous for a virtual exciton created by a $\sigma_+$ photon than by a $\sigma_-$ photon, if the number of (-1/2) Fermi sea electrons is the largest. These carrier exchanges thus differentiate the response to $\sigma_+$ and $\sigma_-$ photons in the case of a polarized Fermi sea. They induce different dielectric constants for the $\sigma_\pm$ parts of the linear probe beam, *i.e.*, different absorptions responsible for the circular dichroism shown in Fig.3(b), and different refractive indices giving rise to the Faraday rotation shown in Fig.3(c).

The theory of this very complicated many-body problem is far beyond the scope of this paper, our old work [17] being rather unsatisfactory: Indeed, it misses the Coulomb repulsion between the Fermi sea electrons and the electron of the photocreated virtual exciton, which is *a priori* as large as the attraction of the photocreated hole.

Let us end by one comment: We could think that absorption with trion formation has close similarity with the formation of an exciton bound to a donor. This is not true: as donors are discernable particles, excitons bound to donors are classical particles; they can be piled up all at the same energy, so that the absorption line barely increases with the donor density. On the opposite, when an exciton is bound to an electron to form a trion, the resulting object is essentially a fermion; an increase in the number of available electrons does not increase the weight of one specific trion line but spreads it over a large energy range: This is conceptually different from an oscillator strength increase.

*c) Experimental set-up.*

The pump-probe experimental set-up used to obtain the photoinduced Faraday rotation spectrum shown in Fig.3(c) is the one of Ref.[11]. It is slightly modified to get the circular dichroism. The pump pulse is circularly polarized and the probe pulse is linearly polarized at 45° from an arbitrary axis (x). The probe light is analyzed after transmission through a quarter-wave plate. The two circular components of this linear probe are then transformed into (x) and (y) polarized lights. The circular dichroism signal, *i.e.*, the variation of ($\alpha_+ - \alpha_-$), where $\alpha_\pm$ is the absorption coefficient of a $\sigma_\pm$ light, is finally obtained using a Glan-laser cube, aligned with the (x) and (y) axes, and a balanced optical bridge [11].

The sample is a one-side modulation-doped CdTe/CdMgTe quantum well, grown by molecular-beam epitaxy on a (100)-oriented GaAs substrate. The well width is 100 Å and the iodine donor layer is at 400 Å from the well. The electron concentration is $1.7 \; 10^{11}$ cm$^{-2}$, as determined by magneto-transmission. The corresponding Fermi energy is 4.1 meV. This rather large doping has been chosen to have a well separated "trion line", as seen from Fig.3(a) which shows the linear transmission of the sample at 2K. The broad line at higher energy is the so-called "exciton line".

The Fermi sea is polarized as described in Ref.[11]: A $\sigma_+$ pump beam creates a certain amount of (+3/2) holes and (-1/2) electrons. Due to a fast hole relaxation, each hole which has flip its momentum recombines with a (+1/2) electron; so that for pump-probe delays larger than the carrier recombination time (80 ps), we end with a spin polarized 2D electron gas, with more (-1/2) than (+1/2) electrons, the damping time of this electronic spin polarization being 190 ps.

Fig.3(b) shows the intensity of the photoinduced circular dichroism signal as a function of the pump and probe photon energy, for a fixed pump-probe delay of 100 ps. The zero signal appears at the "trion energy", E = 1608.1 ± 0.5 meV. Fig.3(c) shows the intensity of the photoinduced Faraday rotation *versus* photon energy, at the same pump-probe delay. This curve has a strong maximum at the same "trion energy". Both spectra show an asymmetric shape with a rather sharp edge on the low-energy side and a broad tail at high energy, in good agreement with the theoretical arguments given above.

Time and spectrally resolved pump-probe experiments have been recently reported in n- or p-doped quantum wells [12,13]. These experiments are performed with short pump-probe delays, so that the transmission of the probe beam is modified by the presence of an exciton and/or trion gas, photogenerated by the pump beam. The experimental results are analyzed within a three-population model, in which interactions between electrons, excitons and trions play an important role. This has to be contrasted with our present work: by performing measurements at a long pump-probe delay, we can study the differential $\sigma_+/\sigma_-$ optical response in the presence of a *single* population, *the polarized electron gas*. This allows to evidence many-body effects on the "trion line" in a quite direct way, through the Fermi sea polarization. This approach offers the opportunity to discuss the nature of the elementary excitations of a polarized electron gas, and their signature in optical experiments.

Studies of the "trion line" with a Fermi sea polarized by a static magnetic field, have also been reported [5,6,10]. In these experiments, the two spin populations have the same Fermi level so that such a polarization does not affect the threshold. On the opposite, our different Fermi levels allow to emphasize the threshold singular behavior of the "trion line".

*As a conclusion*, the so-called "trion line", on the low-energy side of the absorption spectra of doped quantum wells, cannot be due to a trion. Trions are essentially fermions; they cannot be piled up at the same energy but actually spread over a large energy range. Moreover, they do interact with the other carriers of the sample. The low-energy line seen in doped quantum wells, is in fact due to singular many-body effects induced by Coulomb and Pauli scatterings between the photocreated (virtual) exciton and the carriers present in the sample. This is supported by the shape of the circular dichroism spectrum observed with a linear probe beam when the Fermi sea is polarized. This polarization, which allows to differentiate the $\sigma_\pm$ components of the linear probe beam through different carrier exchanges, evidences the dominant role played in this problem by many-body effects resulting from the sudden appearance of a photocreated (virtual) exciton in the Fermi sea.

This work was partially supported by The Research Ministry through an «ACI Jeunes chercheurs 2002» grant, and the «Projet SESAME 2003» n° E.1751.

**FIGURE CAPTION**

**Fig.1 (a)** An electron is taken out of the Fermi sea to make a trion. The Fermi sea reacts to this sudden change by creating electron-hole pairs either across the Fermi level or to fill the hole left in the sea.
**(b)** When the Fermi sea has more (–1/2) than (+1/2) electrons, the minimum energy necessary to make a trion with a virtual exciton made from a $\sigma_-$ photon, is lower than the one made from a $\sigma_+$ photon.
**Fig.2** For the polarized Fermi sea shown in Fig.1(b), the absorption of $\sigma_-$ photons starts at a lower energy than the one of $\sigma_+$ photons. These absorptions are constant in the case of simple trions **(a)**, while they should have a sharp threshold and a long tail **(c)** if many-body effects are included, as explained in the text.
**(b, d)** Difference in the absorption of $\sigma_+$ and $\sigma_-$ photons, deduced from **(a)** and **(c)**.
**Fig.3** Spectra obtained at 2K on a single CdTe doped quantum well having an electron density $n_e = 1.7\ 10^{11}$ cm$^{-2}$. The dots represent the experimental data, while the dashed curves are just guides for the eyes: linear transmission **(a),** photoinduced circular dichroism **(b)** and photoinduced Faraday rotation **(c)** obtained with a polarized electron gas.
**Fig.4** A $\sigma_+$ photon (light wavy line) creates a virtual exciton made with a (+3/2) hole (doted line) and a (-1/2) electron (solid line). **(a)** This exciton can have direct Coulomb interactions (heavy wavy line) with any Fermi sea electron with spin (s= ± 1/2).
**(b),(c)** The photocreated virtual exciton can also exchange its electron with the Fermi sea, in more or less complicated processes. The hole can then recombine with an electron different from the photocreated one.

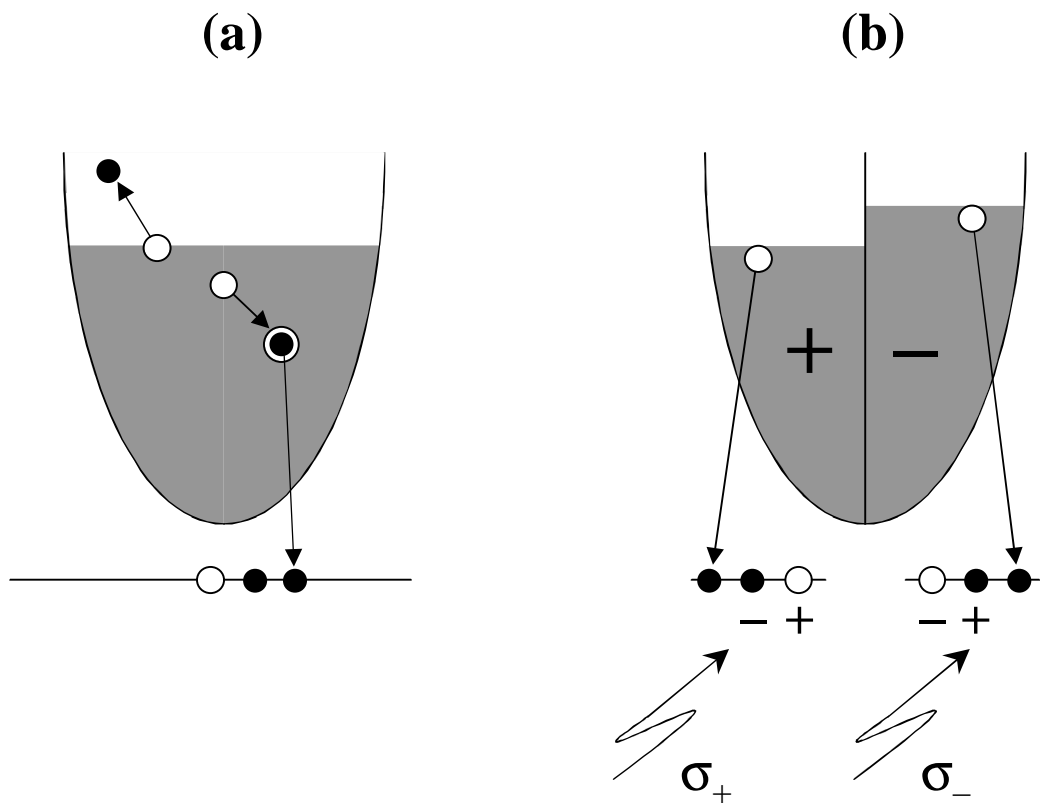

Figure 1

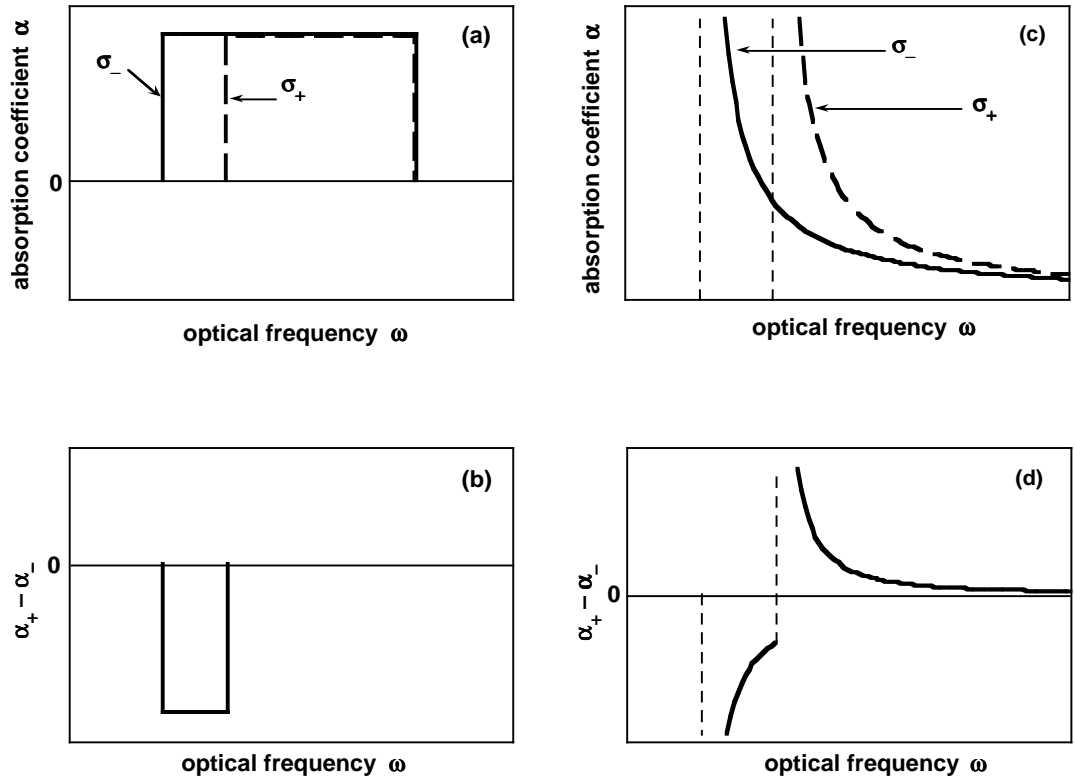

Figure 2

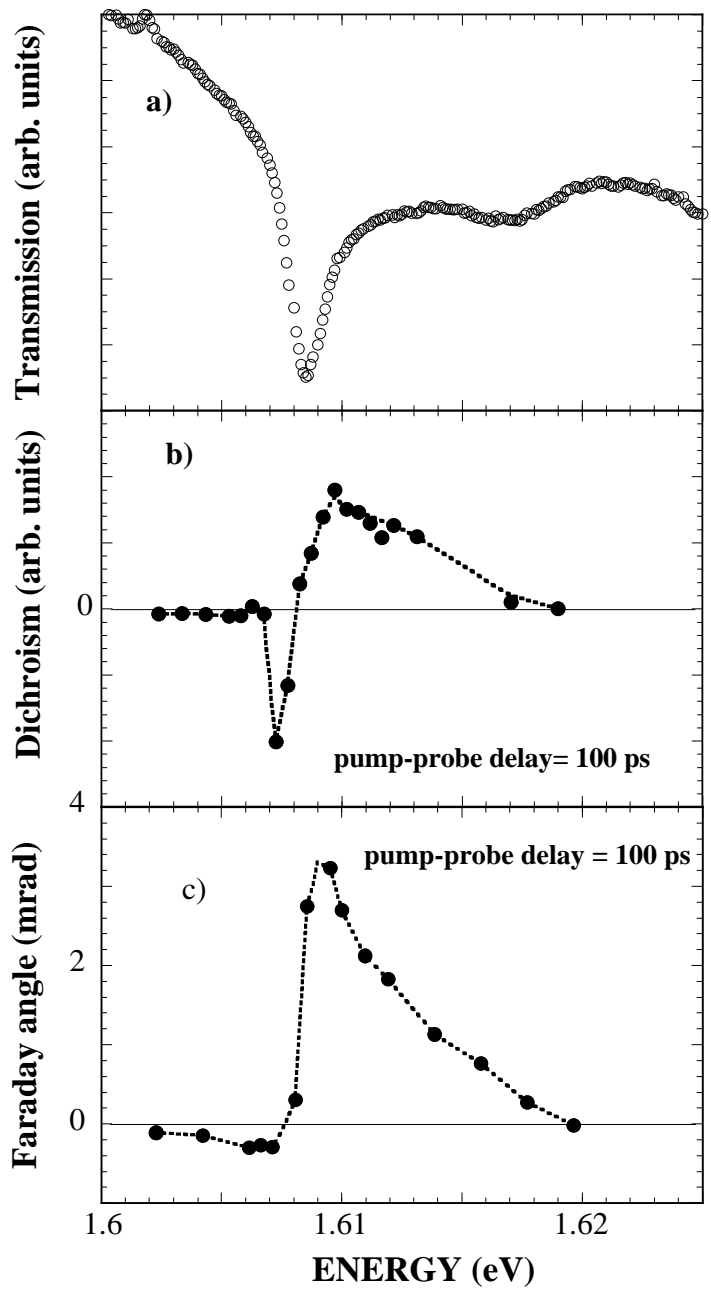

Figure 3

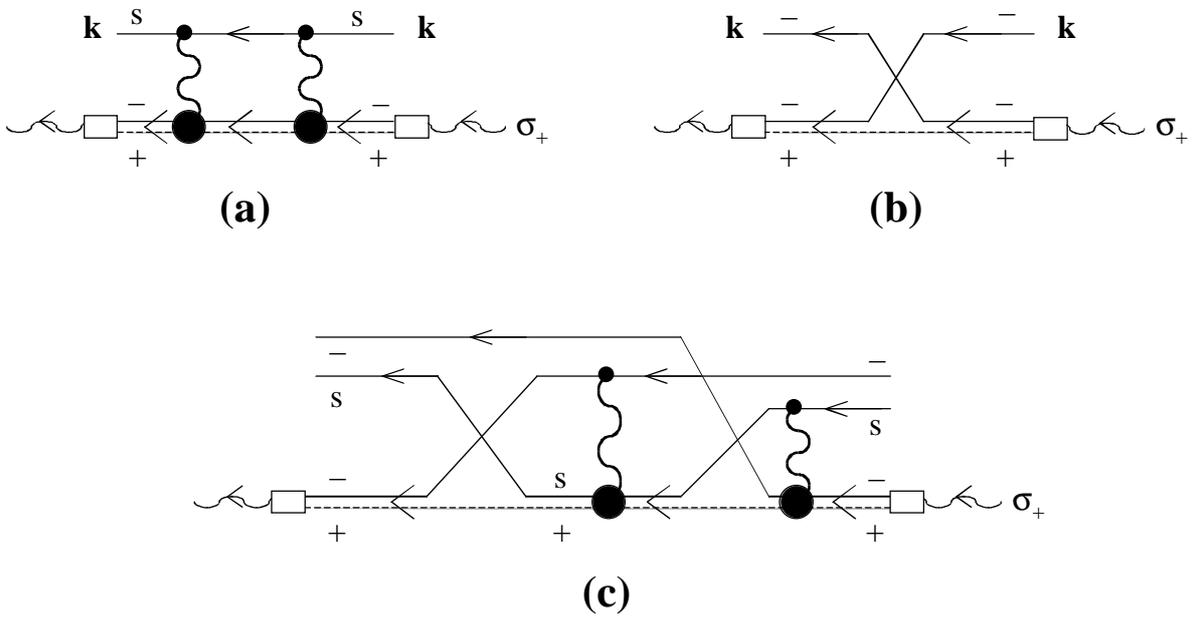

Figure 4